\documentclass[fleqn,usenatbib]{mnras}

\usepackage[T1]{fontenc}

\usepackage{graphicx}	
\usepackage{amsmath}	

\usepackage{subcaption} 

\usepackage{gensymb} 
\usepackage{textcomp}  
\usepackage[normalem]{ulem}
\usepackage[dvipsnames]{xcolor} 
\usepackage{xcolor} 

\newcommand{\red}[1]{\textcolor{red}{#1}}

\title[Hyperbolic limit of bright pulses from Vela]{Hyperbolic limit on the early arrival time of bright pulses from PSR~J0835$-$4510 (Vela)}

\author[A. D. Mahida et al.]{
Angiraben D. Mahida,$^{1,2}$\thanks{E-mail: angiragmahida9@gmail.com}
J. L. Palfreyman,$^{3}$
G. Molera Calvés,$^{3}$
Susmita Sett,$^{4,5}$
\\
$^{1}$School of Physics, University of Western Australia, Crawley WA 6009, Australia;\\
$^{2}$The ARC Center of Excellence for Gravitational Wave Discovery – OzGrav, Australia\\
$^{3}$School of Natural Sciences, University of Tasmania, Private Bag 37, Hobart, Tasmania 7001, Australia\\
$^{4}$International Centre for Radio Astronomy Research, Curtin University, Bentley, WA 6102, Australia\\
$^{5}$CSIRO Astronomy and Space Science, PO Box 76 Epping NSW 1710 Australia}

\date{Accepted XXX. Received YYY; in original form ZZZ}

\pubyear{2023}

\begin{document}
\label{firstpage}
\pagerange{\pageref{firstpage}--\pageref{lastpage}}
\maketitle

\begin{abstract}

Astronomers have studied the Vela pulsar (PSR~J0835$-$4510) for decades. This study analyses almost one hundred hours of single-pulse data collected over three consecutive days from 2016 and 2020. The work investigates the fascinating phenomena of the earlier arrival of brighter pulses with their increase in peak intensity. We found a hyperbolic relation between them by constructing integrated pulse profiles using flux density intervals and examining the relationship between pulse arrival time and intensity. We identified a phase limit of $-0.85~\pm 0.0109$~ms for the earliest arrival of the brightest pulses. This study offers exciting prospects for further exploring the emission regions responsible for the Vela pulsar's regular and giant micro-pulses. 

\end{abstract}

\begin{keywords}
pulsars: individual:Vela
\end{keywords}



\section{Introduction}
\label{sec:intro}

The Vela pulsar (PSR J0835-4510) is an excellent example of studying emission mechanisms due to its normal pulses and giant micro-pulses. This pulsar has been studied for long-term timing analyses \citep{Dodson2007} as well as short-term single pulse studies of normal pulses \citep{Johnston2001}  and giant micro-pulses \citep{Chen2020}. This young, luminous, and close pulsar emits pulses across multiple wavelengths \citep{Pellizzoni2010} and is highly linearly polarised ($> 95\%$ at $1720$~ MHz) with a period of approximately 89.3~ms \citep{Radhakrishnan1969}. Unlike the Crab pulsar, the Vela pulsar does not emit giant pulses exceeding the mean flux density by ten times \citep{Knight2006, Cairns2004, palfreyman2016}. On the contrary, as reported by \cite{Johnston2001}, the Vela pulsar emits giant micro-pulses with high peak-flux density and narrow pulse width at 660 MHz and 1413 MHz. Additionally, consecutive bright pulses with five times the mean flux density of the average pulse were detected, indicating that bright pulses may not be independent random events \citep{Palfreyman2011}. 

We adopted the pulse-to-pulse analysis model introduced by \citet{Krishnamohan1983} (hereafter referred to as KD83) that has provided valuable insights into the nature of the Vela pulsar. The KD83 model involves segmenting adjacent intensity intervals, or "gates", to encompass a wide range of peak pulse intensities. Pulses are then grouped and added based on peak flux density, resulting in gated pulse profiles. The gated profiles revealed that brighter pulses occur earlier in the pulse sequence. The emission region comprises four sub-components representing distinct emission zones at varying distances from the core, as found by KD83. In our study, we aim to extend this model by examining the behaviour of gated profiles at higher peak flux densities and applying it to a larger dataset to explore sub-components' presence. The reasons behind the early arrival of bright pulses remain unexplained in the literature, and the KD83 model has yet to be applied to large datasets like ours. Hence, we examined regular, bright, and giant micro-pulses for six days using the KD83 model and then focused our analysis on a specific dataset. We test the peak flux density against the full width half maxima (FWHM) and phase of the peak to identify any correlations and attempt to discern sub-components in the higher gated profiles.

To establish continuity with today's research while extending the gated profiles by a factor of 20. We have used the term "Fflux-density range(s)" instead of "gate(s)" and "flux-density integrated profile" instead of gated profiles. This paper describes our observations and data reduction process, including a critical comparison to the original analysis conducted by \cite{Krishnamohan1983}, in Section \ref{sec:data}. We then present our results and discuss their implications in Sections \ref{sec:result} and \ref{sec:discuss}, respectively.


\section{OBSERVATION AND DATA REDUCTION}
\label{sec:data}
  
The University of Tasmania's 26~m radio telescope at Mt. Pleasant Observatory near Hobart, Tasmania, Australia, has been used to observe the Vela pulsar for long-term single-pulse studies since $2007$. In this work, we report on 111~hours of the Vela pulsar single pulses recorded in 2016 and 2020. The data for this analysis has a central frequency of $1376$~MHz with a bandwidth of $64$~MHz. Each raw file is $10$~s long, containing around $111$ single pulses. DSPSR (Digital Signal Processing Software for Pulsar Astronomy, \citep{VanStraten2011}) was used to fold the raw data into $8192$ timing bins with a $10.9~\mu$s time resolution. Each file was also integrated in polarisation and frequency using PSRCHIVE \citep{Hotan2004}.

Flux Density calibration was performed by pointing on and off at a known source calibrator and pulsing with a noise diode. Since the L-band setup of the radio telescope rarely changed, these were only performed sporadically.

\begin{figure*}
\center
\includegraphics[ height=75mm]{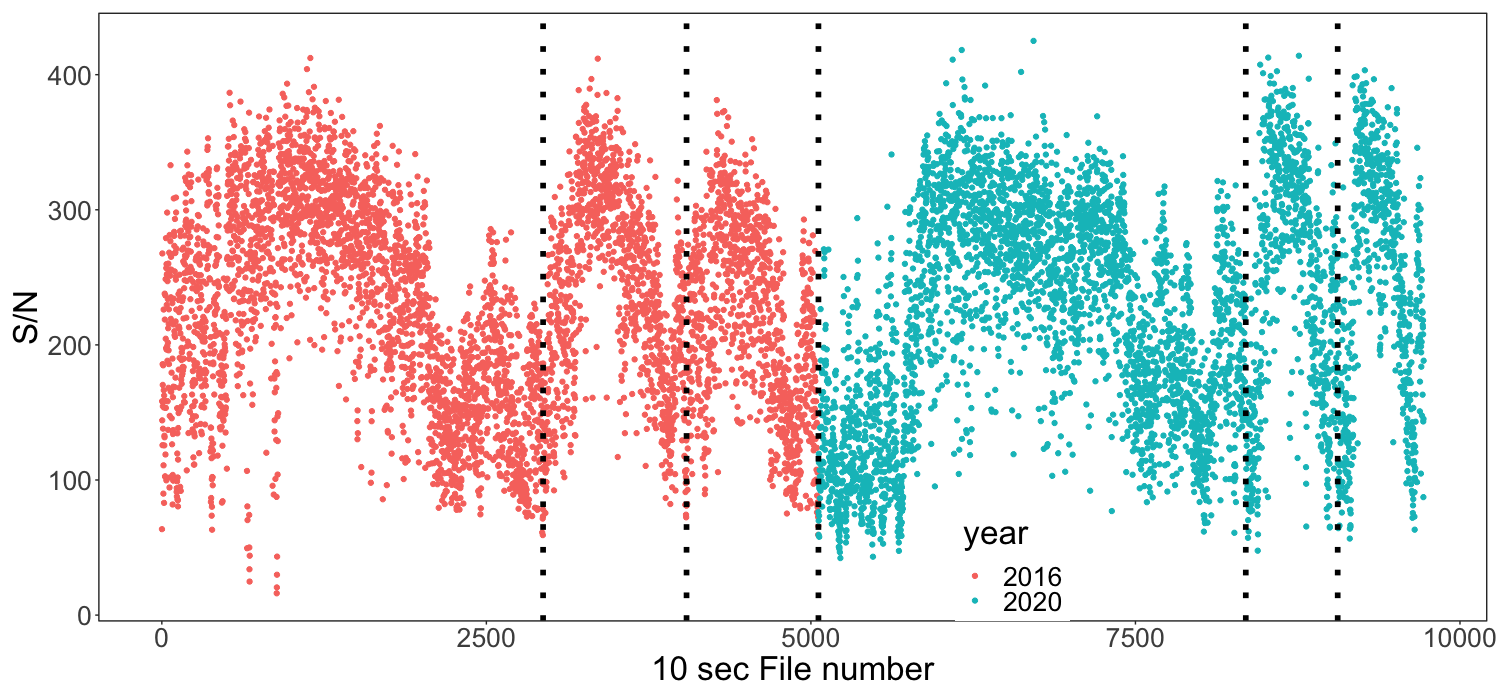}
\caption{Signal-to-noise ratio for each 10-second file. Different colours represent different years, and dotted lines are to separate different days. The X-axis is not linear in time as some daily observations are different lengths, and some data contaminated with RFI has been removed.}
\label{fig:noise}
\end{figure*}

We selected four calibrated observations that were longer than 18 hours and between the selected days in 2016 and 2020. These calibrated days showed that 1.0 on the Mt Pleasant relative scale was equivalent to 63.7~Jy (standard deviation of 0.72~Jy) as shown in Table~\ref{tab:calibration}. For this study, it is important to note that the relative flux densities of the days of observation are consistent. Figure~\ref{fig:noise} shows the signal-to-noise ratio of each 10~s file over the set of selected observations. The noteworthy observation is the apparent presence of cycles. This is most likely due to the differences in sensitivity of the two linear channels and low-noise amplifiers in the L-band receiver at Mt Pleasant. Vela is highly linearly polarised, so the signal moves from channel to channel as Vela passes across the sky. The signal-to-noise values are reasonably consistent across the observation days, even though they are years apart. This also implies that the system temperature was fairly stable throughout our analyses.

\begin{table}
    \caption{Results of four days (each longer than 18 hours) in 2017 where flux density calibration occurred. The mean value for the 1.0 reference level is 63.7~Jy.}
    \label{tab:calibration}
    \centering
    \begin{tabular}{cccc}
    \hline
    Date & Length (h) & 1.0 Ref (Jy)\\
    \hline
    2017\_036	& 18.77	& 63.9 \\
    2017\_040	& 18.66	& 64.1 \\
    2017\_071	& 18.63	& 64.1 \\
    2017\_115	& 18.77	& 62.6 \\
    \hline
    \end{tabular}
\end{table}

\cite{palfreyman2016} showed the cyclic nature of the FWHM of the integrated pulse profile of a day, which could be related to the amount and type of pulses emitted by the Vela pulsar. $1.52$~ms was the median value of the extensive dataset \citep{palfreyman2016}. Considering this possible relation, we selected a day, Modified Julian Date (MJD) $58851$, where the FWHM was greater than $1.52$~ms (hereafter 2020-data) and another day where it was less than $1.52$~ms, MJD $57661$ (hereafter 2016-data). The FWHM of the integrated pulse profiles of 2020-data and 2016-data were around $1.522$ and $1.499$~ms, respectively. This work assumes that MJD $58851$ (from 2020) could be a high-activity day, and MJD $57661$ (from 2016) could be a less active day, expecting fewer bright and giant micro-pulses. Bright and giant micro-pulses have higher peak flux density and smaller FWHM. The results of these two different activity phases of the Vela pulsar are discussed later in Section \ref{sec:result}. Another influential factor was choosing days with minimum apparent radio frequency interference (RFI). Observed data of pulsar radio emission often suffer from various types of RFI, which could be caused by human communications technologies such as satellites, mobile base stations, or navigation radars. We did not select a day with an open day at the observatory.  

We followed the model created by KD83 but with minor changes in the interval to further categorise and analyse our data. The intervals of peak flux for the KD83 gated profiles were not evenly spaced, but we distributed our flux-density range (hereafter FD-range) with smaller and equal intervals. We have created 80 peak flux intervals for the distribution of pulses, each of our FD-range being $1$~unit wide. For example, FD-range $1$ would include pulses of peak flux $1.00\leq S_p<1.99$, in arbitrary units (AU); i.e., $63.7\leq S_p<126.76$ Jy.

We integrated every pulse from each FD-range to create flux-density integrated (hereafter FD-integrated)  profiles. In other words, we were averaging profiles within the same energy range. These FD-integrated profiles had signal-to-noise ratios varying from as high as 9075 (FD range 3 with 169942 pulses) to as low as 77 (FD range 76 with one pulse) because of a varying number of extracted pulses. We added pulses from the other two consecutive days to improve noisy FD-integrated profiles (mostly higher (brighter) profiles; > 20 for 2016-data and > 30 for 2020-data) with lower SNR values. We manually checked each pulse from brighter profiles to remove radio frequency interference (RFI). So all $16944$ pulses in these higher FD-range were manually checked and removed if RFI was predominant. In this analysis, we had $4,358,856$ pulses, and $1,348,242$ pulses were used to create our FD-integrated profiles. The pulse number difference is because the lower FD-ranges have pulses from just one day, and higher FD-ranges from three days. After comparing the number of pulses of 80 FD-ranges from both activity days, we focused on the phase of the peak of FD-integrated pulse profiles from high activity days to analyse the nature of the arrival time and FWHM. The sub-components are also analysed from a brighter profile of the 2020-data. 

The peak flux density ($S_p$) of bright profiles was calculated using the radiometer equation, 

\begin{equation}
S_p=S / N  \frac{S_{sys}}  {\sqrt{n_p  \times \Delta f  \times  T_{i n t}}}
\label{eq:sp}
\end{equation}

where $ S_{sys} = 470~Jy $ is the system equivalent flux density \footnote{\url{https://ra-wiki.phys.utas.edu.au/}}, $n_p = 2 $ is the number of polarisation summed, $\Delta f =$~64~MHz is the effective bandwidth and $T_{i n t} = 10.9~\micro$s is the sampling interval (\cite{LKbook}).


\section{RESULTS}
\label{sec:result}

The peak flux of the integrated pulse profile of $\sim18$~hr from 2020-data, at $1376$~MHz is $106.29 ~(\pm 0.097)$~Jy arriving at phase $-0.04 ~(\pm 0.01) $~ms and is similar for 2016-data. The uncertainty associated with phase value is from bin width. As mentioned before, the FWHM of $\sim 18$~hr integrated profile from 2020-data is $1.522 ~(\pm 2.0 \times 10^{-04})$~ms whereas, it is  $1.499 ~(\pm 2.156 \times 10^{-04})$~ms for 2016-data. The mean flux density (MFD) of integrated pulse profiles from both datasets is $2.3596 ~(\pm 2.6 \times 10^{-04})$~Jy. According to the relationship described in \cite{palfreyman2016} between the width and Vela's activity, the data from 2020 corresponds to high activity days of the Vela pulsar, while the data from 2016 corresponds to low activity days. 

\subsection{Pulses and their distributions}

Following the definition provided by \cite{Palfreyman2011}, pulses exhibiting integrated flux density at least five times greater than the average pulse flux can be classified as "bright pulses". The term "giant micro-pulses," as defined by \cite{Johnston2001}, refers to a cluster of pulses with widths ranging from 50 to 400 microseconds. At a frequency of 1413 MHz, the brightest pulse recorded by \cite{Johnston2001} had a peak flux density exceeding 2500 Jy, more than 40 times the peak flux density of an integrated pulse profile. These two criteria prove useful for approximately distributing our gates.

FD-range with an index greater than or equal to 20 exhibit integrated profiles that possess five times the mean flux density of a daily integrated profile, with a FWHM value of no more than $0.536$~ms. The pulses integrated within these profiles have a peak flux of at least 1275 Jy. Meanwhile, individual pulses extracted from FD-range 47 exhibit a peak flux density exceeding 3000 Jy. Based on these characteristics, we have categorized our pulses into three types (normal pulses, bright pulses, and giant micro-pulses). Consequently, FD-ranges zero to 19 contain normal pulses, FD-ranges 20 to 46 contain bright pulses, and FD-ranges 47 onwards consist of FD-integrated profiles featuring giant micro-pulses. The total counts of each pulse type for both data sets are provided in Table \ref{tab:pulses}.

\begin{table} 

    \caption{The table presents the number of pulses for two datasets. The first two columns display the Flux Density (FD) range and the corresponding pulse type integrated within each range. The last two columns provide the pulse count within each FD range and the average pulses per hour for that FD-range.}
    \label{tab:pulses}
    \centering
    \begin{tabular}{cccc}
    \hline
    Range  & \red{Pulse} Type & \red{2020-data (hr$^{-1}$)} & \red{2016-data (hr$^{-1}$)}\\
    \hline
    00 - 19  &  Normal  & 656925 (1892.4)   &  680544 (2048.7)\\
    20 - 46  & Bright  &  2921 (3.55) & 1965 (1.33)\\
    47 - 79  &  Giant Micro &  360 (0.2053) & 132 (0.0747) \\
    \hline
    \end{tabular}
\end{table}

\begin{figure}
\center
\includegraphics[width=85mm]{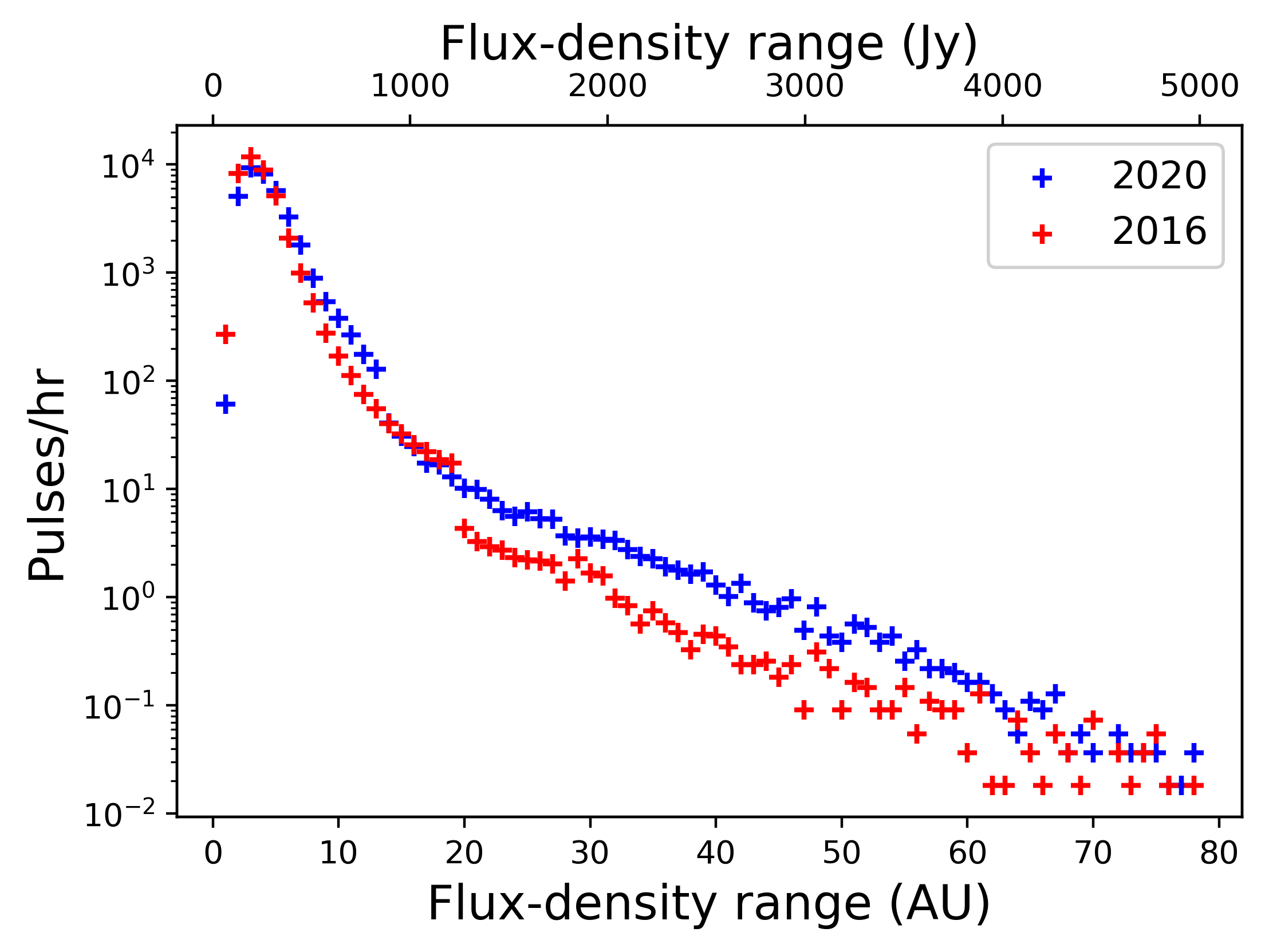}
\caption{Comparison of the number of pulses per hour for each flux density integrated profile. The red crosses correspond to pulses recorded on low-activity days, the 2016-data. The blue crosses represent high-activity days, the 2020-data. The FD-range is shown in Arbitrary Units (AU) and in Jansky (Jy). Pulses contaminated with RFI have been removed from FD-range 14 and onwards for the 2020 data, and 13 and onwards for the 2016 data.}
\label{fig:pulses}
\end{figure}

\begin{figure}
\center
\includegraphics[width=85mm]{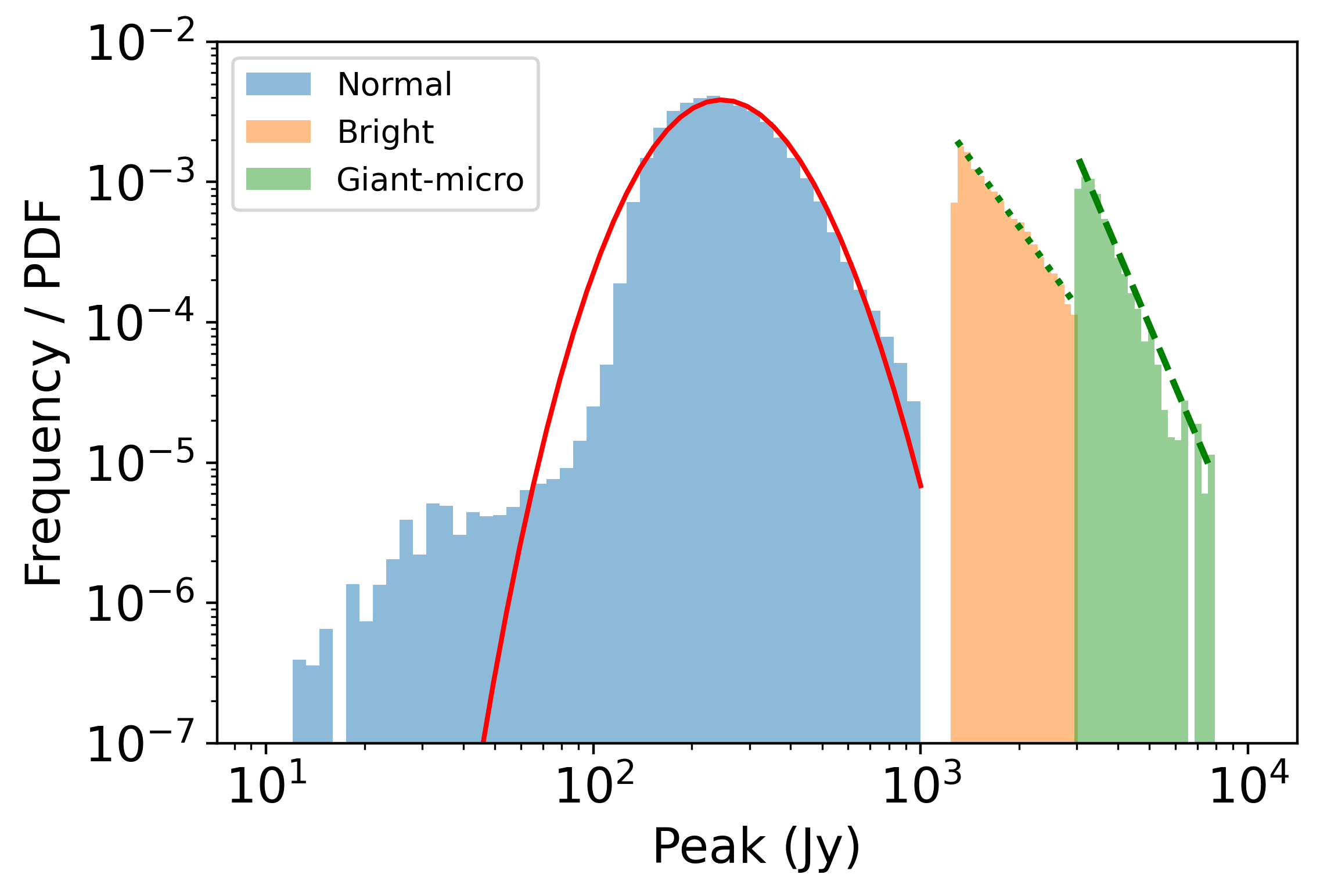}
\caption{PDF and histogram of the peak-flux densities for the normal pulses (blue), bright pulses (orange) and detected giant micropulses (green) from the Vela pulsar. The solid red line shows the expected logarithmic normal distribution, which fits the PDF for the normal pulses well. While the PDF for the bright and giant micropulses are described by power-law distributions (green-dashed and green-dotted lines, respectively).) }
\label{fig:hist}
\end{figure}

 The comparison of the total number of pulses for both data sets has been tabulated in Table~\ref{tab:pulses}. Each FD-range corresponds to a different duration; for example, FD-range 1 to 29 from 2020-data have pulses from $\sim$ 18 hours, whereas FD-range 30 onwards pulses are from around 54 hours and 48 minutes. For 2016-data, pulses up to FD-range 19 are from nearly 17.5 hours, whereas the pulses of brighter profiles (FD-range 20+) are from 55 hours and 15 minutes. We summarise this by showing pulses per hour for each pulse type. For 2020-data, when the Vela pulsar was more active, high-activity days recorded more bright pulses and giant micro-pulses than low-activity days (2016-data). From the analysis of around 111 hours spanned over six days from two different years, we have received almost double the amount of giant micro-pulses and bright pulses for high-activity days than low activity days. As expected, the number of normal pulses is higher for 2016-data. As seen in Figure \ref{fig:pulses}, the number of pulses is constantly high for 2020-data from FD-range five onwards. The disparity observed in the plot near FD-ranges 14 (2020-data) and 20 (2016-data) can be attributed to the differences in the duration of analysis and RFI removal. Specifically, for the 2016-data, pulses were extracted from three days ($\sim$ 55 hours 15 minutes) for FD-range 20 and above. In contrast, for the 2020-data, pulses for FD-ranges greater than 30 were extracted from three days ($\sim$ 54 hours 48 minutes), as shown in Figure \ref{fig:pulses}. In addition, We have manually removed RFI by inspecting individual pulses from FD-range 14 and onwards for 2020 data, and 13 and onwards for the 2016 data. For these RFI removals, $972$ (12 per cent of checked pulses) and $2172$ (23 per cent of checked pulses) pulses have been removed from 2020-data and 2016-data, respectively. The number of pulses per hour plotted in this figure is in log-scale to cover a broader range of data. The FD-range with the highest pulses extracted for 2020-data and 2016-data are 169942 and 206451, respectively. The lowest number of pulses we found for FD-integrated profiles in both datasets is one or zero.

It's worth noting that during the extraction of FD-range 0, we observed 119 pulses in the 2016-data, but none in the 2020 data. Given that the days in question were characterised by high activity, we anticipated a higher number of pulses in the upper FD-range, which makes the absence of any pulses in FD-range zero supportive evidence for this hypothesis. Figure \ref{fig:hist} focuses only on 2020-data and represents three histograms for three types of pulses along with the Probability Density Function (PDF) of the peak-flux; normal pulses (blue), bright pulses (orange) and giant micro-pulses (green). As the plot shows, the normal pulses follow a logarithmic normal distribution (red solid line) similar to \cite{Johnston2001,Kramer2002,Chen2020}. The bright (green-dotted line) and giant micro-pulses (green-dashed lines) follow the power-law distribution. The Probability Density Function is useful for distinguishing between different types of pulses. Since the PDF of bright pulses has not been analyzed separately in previous studies, it was unclear what type of distribution to expect. However, after evaluating various possible fits, a power-law distribution was determined to be the best fit based on the adjusted R-square value of $0.99059$.

\subsection{Gated profiles}


The pulses from our FD-range zero would be equivalent to pulses from the first 10 KD83 gated profiles, but we have 119 pulses recorded with peak flux density $0 - 63.7$~Jy for FD-range zero, only from 2016-data. In KD83, the gated profiles have 87040 pulses analysed from 2.2 hours of data. The highest gate, 15, has pulses with a peak flux density greater than $156\,$Jy. However, KD83 has not mentioned the upper limit of the peak flux density interval of gate 15 or the peak of the brightest pulse they found \citep{Krishnamohan1983}.

\begin{figure*}
\center
\includegraphics[width = 190mm]{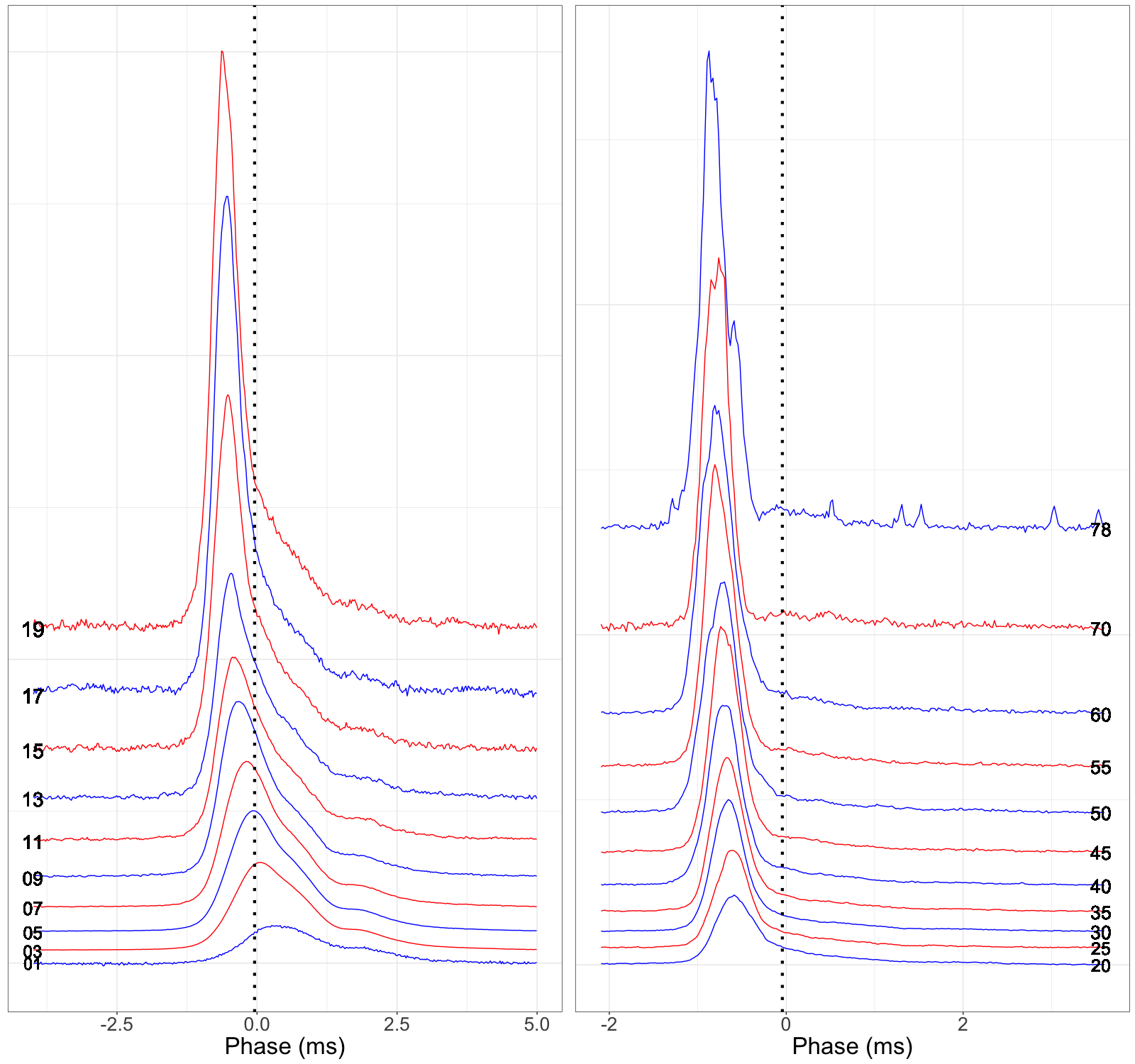}
\caption{Comparison of lower (fainter) and higher (brighter) FD-integrated profiles (2020-data). The left panel displays selected fainter profiles up to FD-range 19, while the right panel presents selected brighter profiles. The FD-range numbers are indicated on the edges of the profiles. The black dotted vertical line represents the phase value of the integrated pulse profiles of $\sim 18$~hr, included for comparison purposes. It is important to note that any instances of radio frequency interference (RFI) have been manually removed from FD-ranges 14 and higher. The scaling of the Y-axes on these panels is arbitrary (although consistent within each panel). The vertical spacing between the profiles is selected for viewing. This is similar to the idea of KD83's Figure 5 to show the change in phase and shape. }
\label{fig:gates}
\end{figure*} 

Nevertheless, with a rough approximation, our first four FD-integrated profiles have pulses up to $317\,$Jy, including all fifteen KD83 gated profiles. The total number of pulses we have analysed ($1,362,042$), with the highest peak flux of $5076\,$Jy has been recorded for FD-range 79 from 2016-data, which is 48 times the peak flux density of the integrated pulse profile of $\sim 18$~hr. The $\sim 18$~hr is the maximum period when the Vela pulsar is observable at Mt Pleasant in a day and that is a good baseline for comparison. The other noteworthy difference is that our data was recorded at a centre frequency of  $1376\,$MHz whereas KD83 gated profiles have been recorded at $2295\,$MHz. The width of our flux-density intervals is uniform, covering six days from two different years, approximately 111 hours.

Figure~\ref{fig:gates} shows the selected FD-integrated profiles from our 2020-data. The left panel has fainter profiles (01-19), whereas the right panel has brighter profiles (20 and up) where RFI has been manually removed for FD-range 14 and beyond. The shifting of phase is noticeable for fainter profiles. The edges show the FD-range numbers; alternative colours have been selected to differentiate profiles. While our data are folded with 8192 phase bins (resulting in a time resolution of 10.9 $\mu$s), the FD-integrated profiles were analysed with a reduced resolution of 4096 phase bins in order to lower the noise in the profiles.

\begin{figure}
\center
\includegraphics[width=85mm]{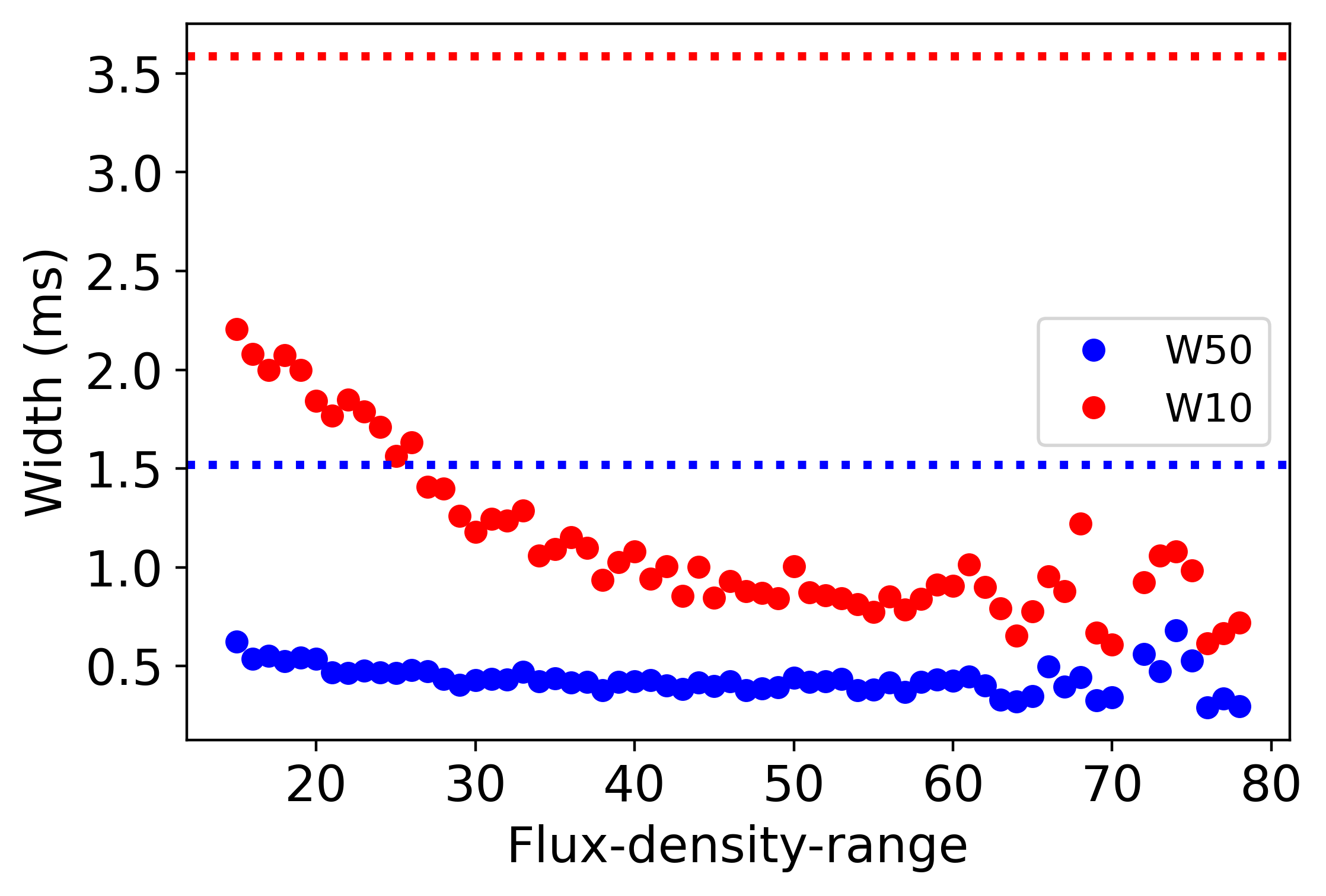}
\caption{Profile widths at the 50 and 10 per cent levels, highlighting the contrasting characteristics between fainter profiles (normal pulses) and brighter profiles (bright and giant micro-pulses). The horizontal lines represent integrated $\sim 18$~hr profile widths. This plot only has FD-integrated profiles where pulses contaminated with RFI were completely removed.}
\label{fig:fwhm}
\end{figure}

\begin{figure}
\center
\includegraphics[width=85mm]{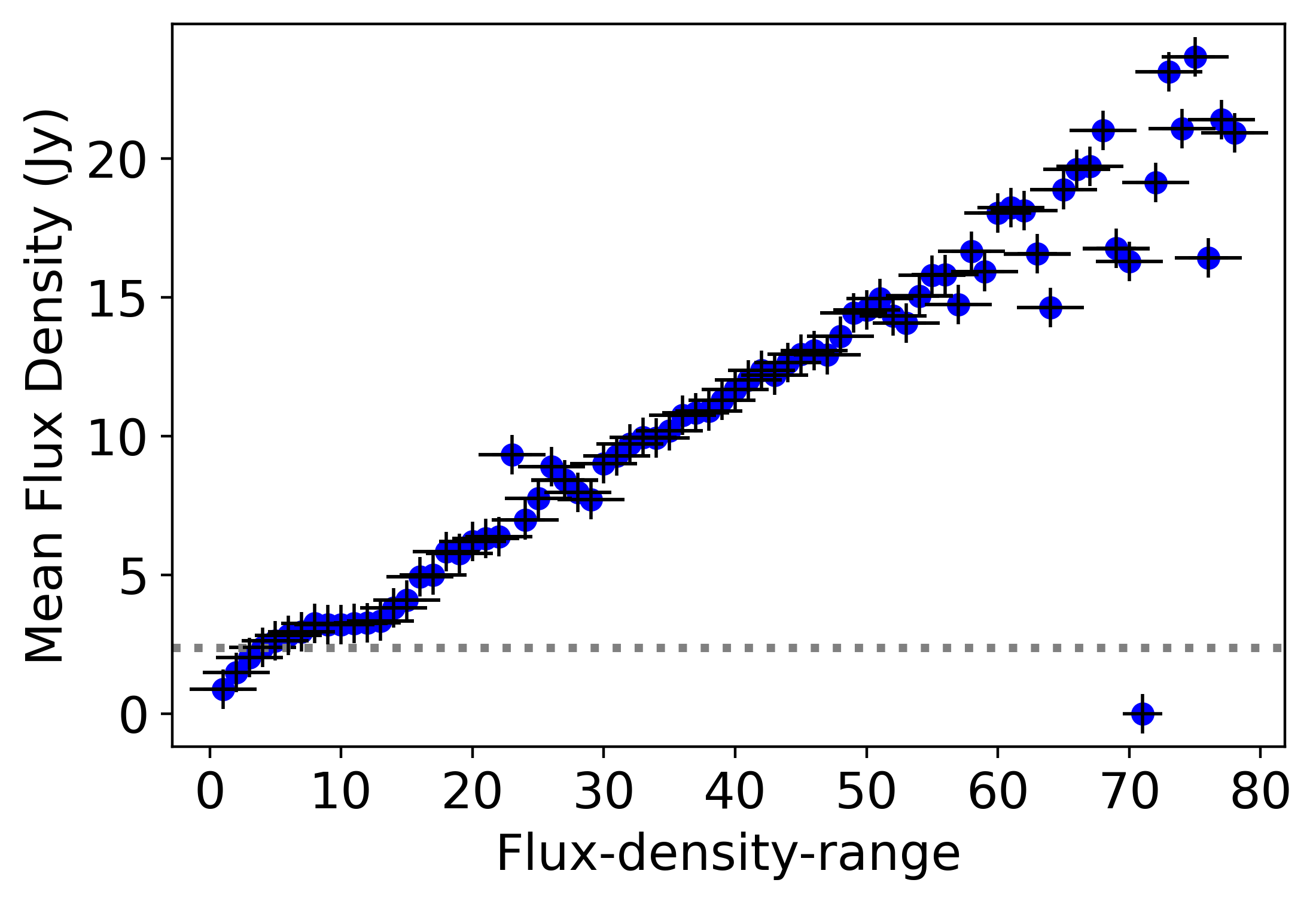}
\caption{The plot displays the mean flux density of the FD-integrated profiles, accompanied by a horizontal grey line representing the mean flux density of the integrated pulse profile of $\sim 18$~hr. Note that we haven't found any pulse for FD-range 71.}
\label{fig:mfd}
\end{figure}

\begin{figure}
\center
\includegraphics[width=85mm]{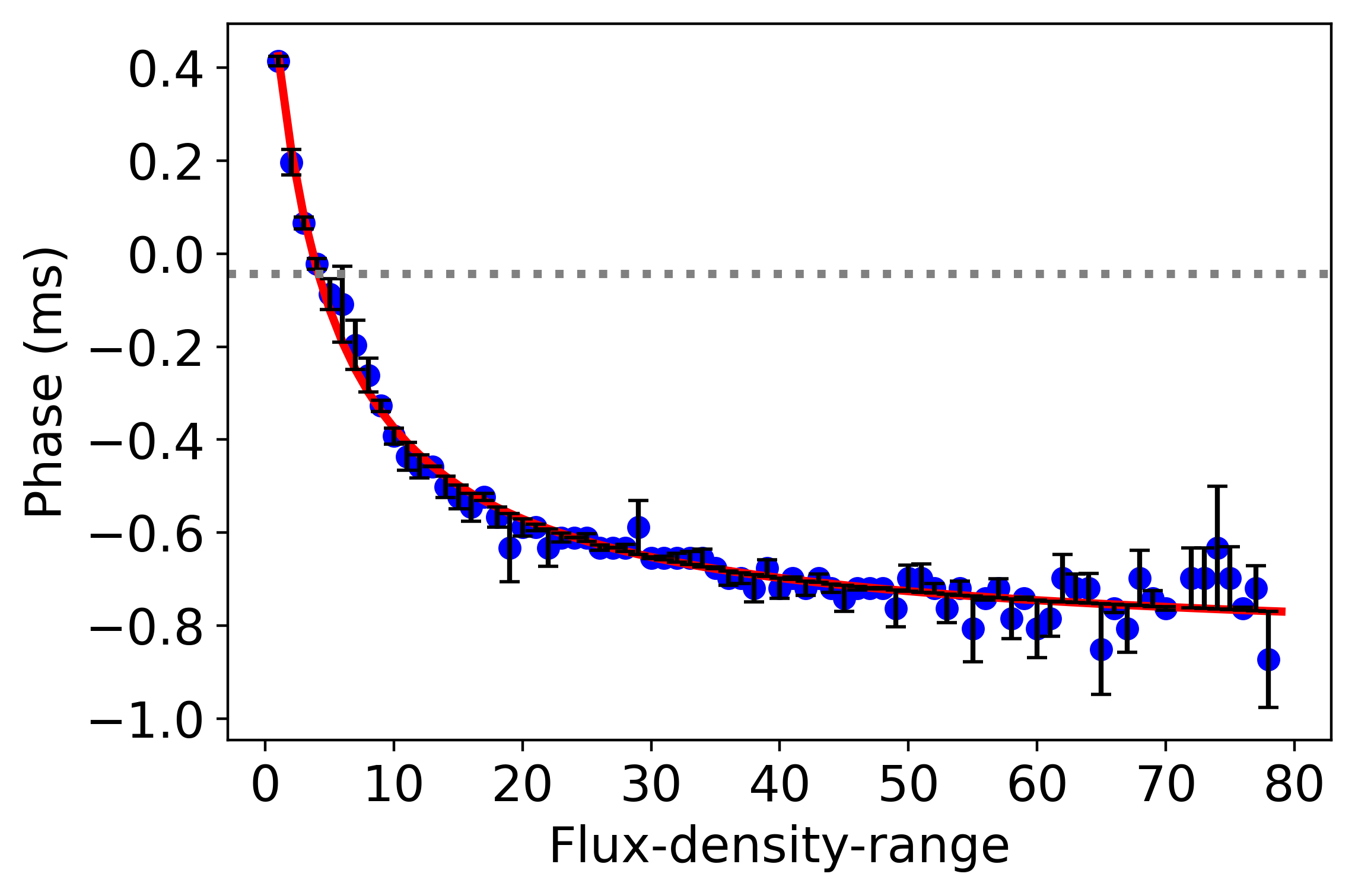}
    \caption{The figure depicts the relationship between the peak phase (ms) and the corresponding FD-range. The horizontal grey line represents the phase of the peak of the integrated pulse profile of $\sim 18$~hr. The circles represent the relative arrival time of each FD-integrated profile, while the red line represents the fitted hyperbola curve. The correlation coefficient for the fit is determined to be 0.9743. The fit function utilized in this analysis is expressed as $ f(x) = \frac{6.852} {x - 4.37} - 0.8517 $,where x is the flux density range.} 
\label{fig:phase}
\end{figure}

We also analysed the FWHM, mean flux density (MFD) and phase of the peak flux of our FD-integrated profiles for 2020-data shown in Figure~\ref{fig:fwhm}, \ref{fig:mfd}, and \ref{fig:phase} respectively. The range of flux densities covered by our FD-ranges goes well beyond that used by KD83, allowing us to study the nature of pulse profiles with much higher peak densities. The nature of fainter profiles is well understood \citep{Krishnamohan1983}, but unexpected behaviour was noticed for brighter profiles. The width at 50 (Blue) and 10 (Red) per cent levels are plotted in Figure \ref{fig:fwhm} along with the horizontal lines representing the widths for an integrated profile of  $\sim 18$~hr. The FWHM of fainter profiles is distributed across broader regions. In contrast, the majority of brighter profiles exhibit approximately similar values. The pulses in FD-range 14 and beyond are manually verified to remove RFI, hence the sudden drop in the width plot. The brighter profiles are scattered, and we can see the difference after FD-range 20. The MFD is plotted in figure \ref{fig:mfd} with the horizontal grey line showing the MFD of the integrated profile of $\sim 18$~hr. We have FD-range 79 with almost nine times MFD.

Since it is well known in the literature that the bright pulses arrive earlier, we compared the phase values of the peak of our FD-integrated profiles. The graphical illustration of the nature of phase values for the peak is shown in Figure~\ref{fig:phase}. Using MATLAB, the best fit we could find for this plot is hyperbolic. The fit equation is
\begin{equation*}
 \Delta = \frac{6.852} {S_p - 4.37} - 0.8517  
\end{equation*}
The best-fit coefficients and their confidence bound for the fit function $ f(x) = \frac{a}{x-b}+c $ are $a=6.852~(6.012, 7.693), b=-4.37~(-5.126,-3.613), c=-0.8517~(-0.8726, -0.8308)$. The sum of squares due to error is $0.1136$ and the adjusted R-square value of the fit is $0.9729$. A hyperbolic fit suggests the arrival times of the FD-integrated profiles asymptote to a value $850$\,~microseconds earlier than the fiducial centre of the pulse.

\subsection{Emission zones and giant micro-pulses}

\begin{figure*}
\centering
\begin{subfigure}[b]{0.45\textwidth}
  \includegraphics[width=\linewidth, height=65mm]{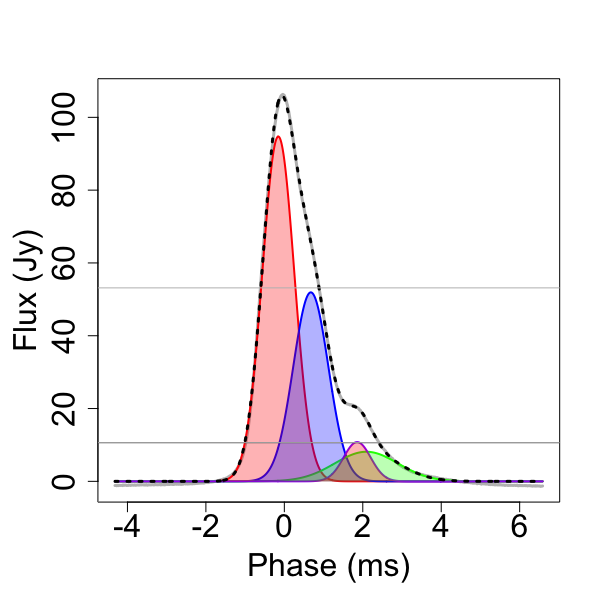}
  \caption{}\label{fig:subcom}
\end{subfigure}
\begin{subfigure}[b]{0.45\textwidth}
  \includegraphics[width=\linewidth, height=65mm]{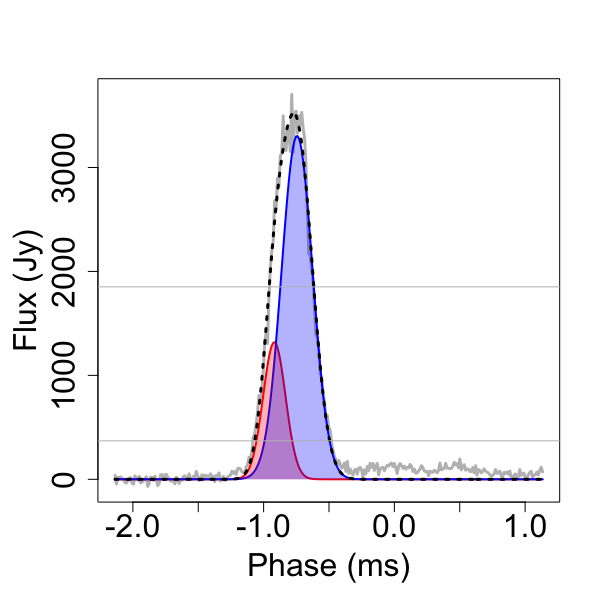}
  \caption{}\label{fig:gate70}
\end{subfigure}
\caption{(a) The integrated pulse profile of $\sim 18\,$h (on MJD 58851) at $1376\,$MHz broken into four Gaussian components. The integrated pulse profile is a grey line overlapped with a black dotted line representing the sum of the four components. The horizontal grey lines show the 10 and 50 per cent levels of the peak flux of the integrated pulse. (b) The FD-integrated profile of FD-range 70 is broken into two Gaussian components, red and blue lines. The observed profile is a grey line overlapped with a black dotted line representing the sum of the two components. The horizontal grey lines show the 10 and 50 per cent levels of the peak flux of the integrated profile.}\label{fig:subcomp}
\end{figure*}

The analysis by KD83 concluded that the emission from the Vela pulsar has four different emission zones with different widths that are located at different distances from the neutron star surface. We attempt to find Gaussian sub-components for our integrated profile of day MJD~58851 at $1376\,$MHz, as shown in Figure \ref{fig:subcom}. The sum of these components has been plotted with a black dotted line, and the original recorded integrated pulse profile is in grey. The resulting two overlapping lines support our attempt to find sub-components. The horizontal grey lines show the 10 and 50 per cent levels of the peak flux density of the integrated pulse profile. We find two similar Gaussian sub-components for one of the brighter profiles of 2020-data, FD-range 70, Figure \ref{fig:gate70}. As shown in the figure, the FD-integrated profile is plotted in black lines with two sub-components in red and blue. This plotting determined how those four components behave at brighter profiles, and the results would suggest two possibilities. There are only two emission zones for brighter pulses; the others are too small to observe, following \cite{Krishnamohan1983}. Notably, the phase value of the peak for these sub-components of FD-integrated profile 70 is different than the integrated profile sub-components.

\begin{figure}
\center
\includegraphics[width=90mm]{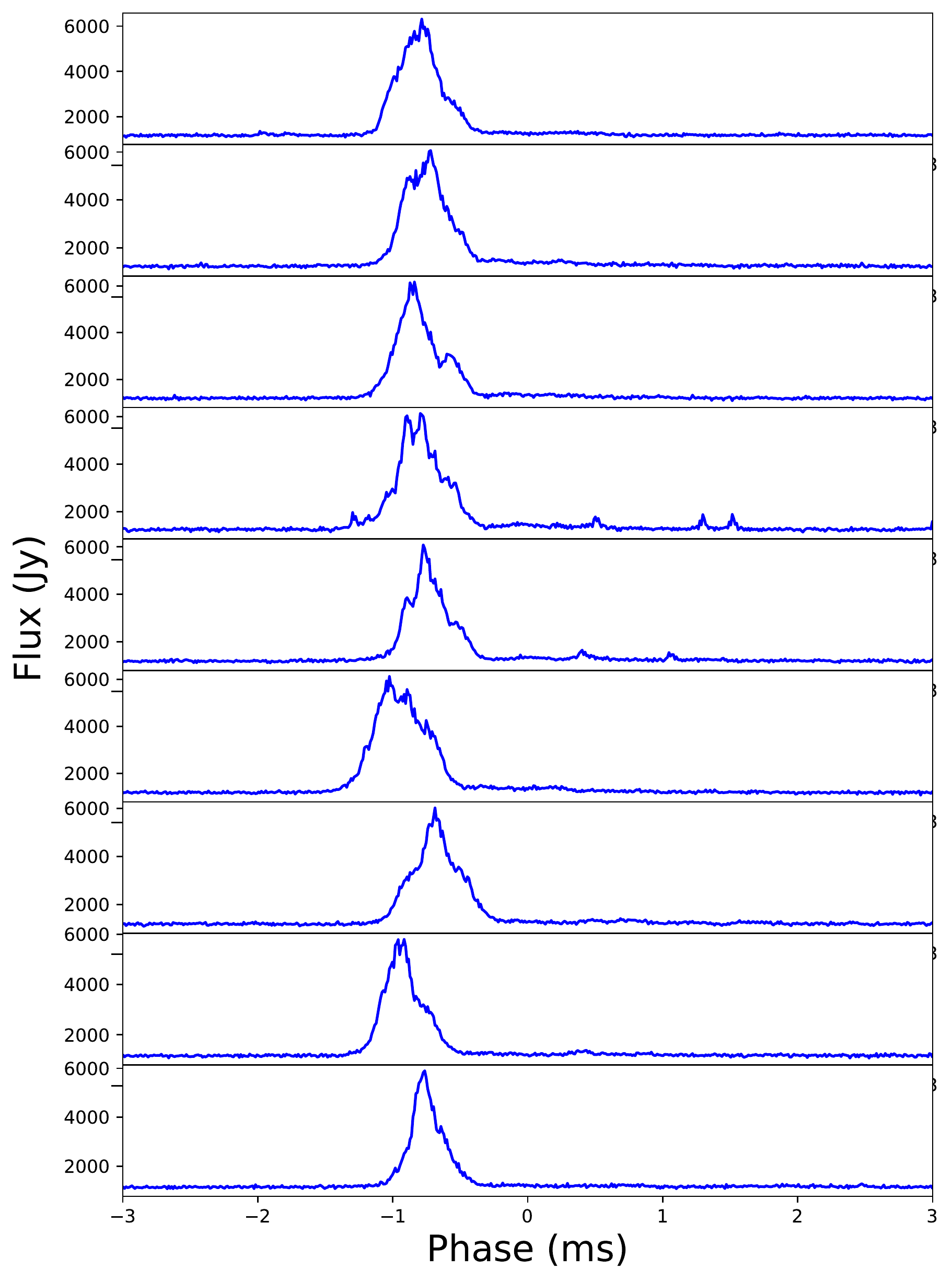}
\caption{The nine brightest pulses detected at $1376$~MHz. These pulses have $8192$ timing bins with a $10.9~\mu$s time resolution.}
\label{fig:nine}
\end{figure}

\begin{figure}
\center
\includegraphics[width=85mm,height=60mm]{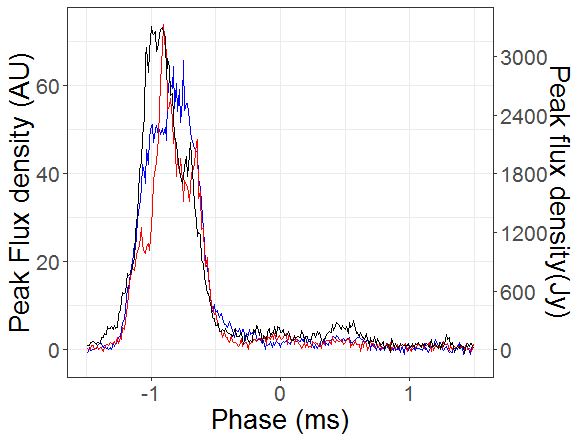}
\caption{Three giant micro-pulses having micro-structure are plotted in the above figure. The X-axis shows the phase values, whereas the Y-axes represent peak flux densities in arbitrary units (left) and Jy (right). The three different colours indicate three different giant micro-pulses.}
\label{fig:micro}
\end{figure}

\begin{figure}
\center
\includegraphics[width=85mm]{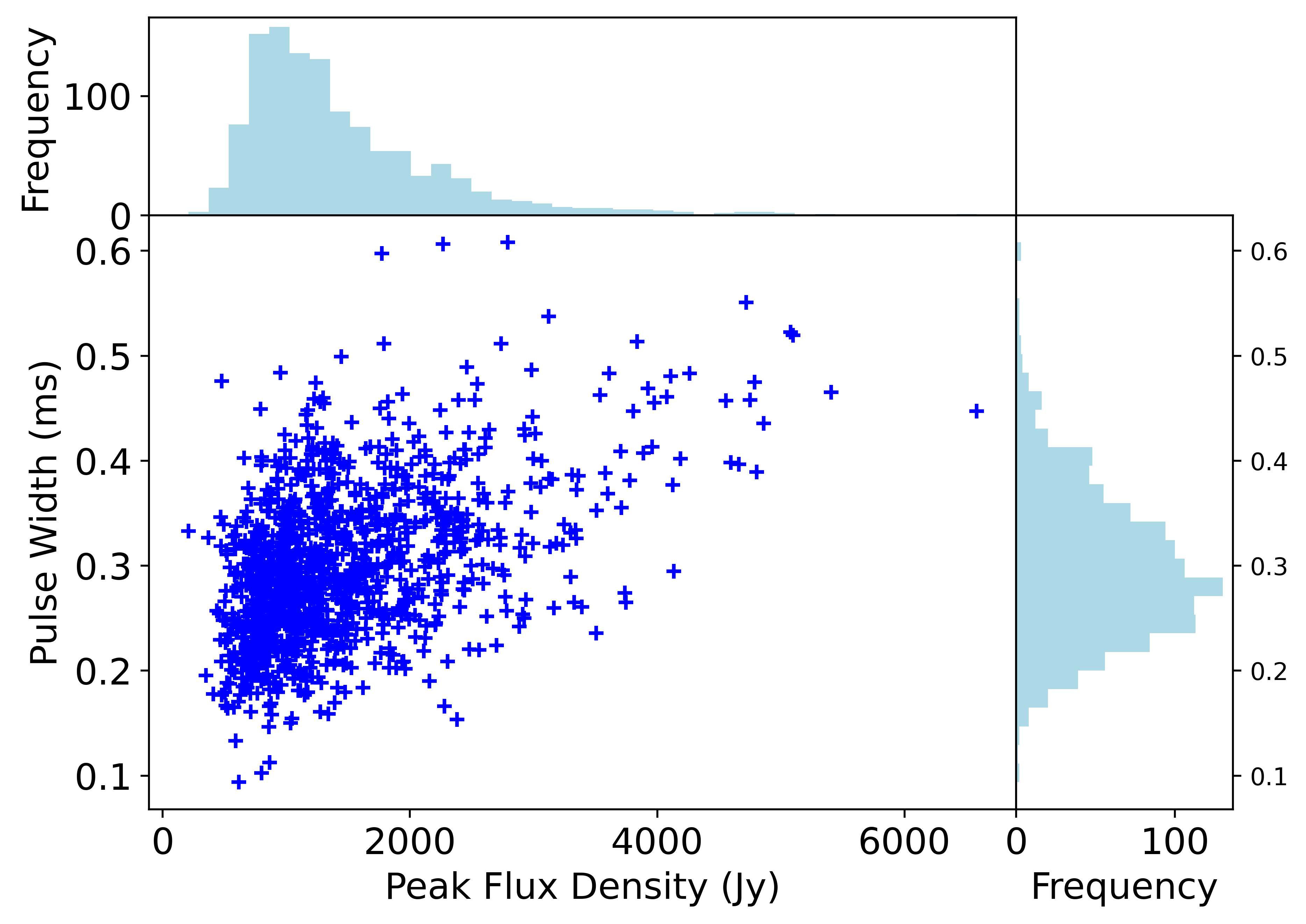}
\caption{Main panel: Scatter plot of width and peak flux density for pulses with peak flux more than $2500$~Jy. Top panel: Histogram of peak flux densities. Right panel: Histogram of pulse widths.}
\label{fig:bright}
\end{figure}

Our analysis captured 986 pulses with peak fluxes exceeding 2500 Jy. Among these pulses, 490 exhibited peak fluxes surpassing 3000 Jy, while 83 pulses displayed peak fluxes exceeding 4000 Jy, which are particularly noteworthy. Figure \ref{fig:nine} showcases the nine brightest pulses from our analysis, illustrating their distinct temporal structures. As expected, these pulses arrive earlier than normal pulses. Multiple peaks and varying pulse widths can be observed in these pulses, such as the fourth panel from the top in Figure \ref{fig:nine}, indicating their complex nature.

In contrast to occasional bumps at the trailing edge of the pulse profile reported by \citet{Johnston2001}, our analysis reveals the presence of small bumps near the peak in some pulses. \citet{Johnston2001} initially detected micro-structure in the Vela pulsar at 660 MHz and was later confirmed at different frequencies by ~\citet{Kramer2002}. We have identified a similar microstructure, but not as wide as the original microstructure reported by \citet{Johnston2001}. Figure \ref{fig:micro} displays three giant micro-pulses' pulse profiles, illustrating the microstructure's presence. The blue and black lines represent pulse profiles of giant micro-pulses from the 2020 dataset (FD-range 66 and 73, respectively), while the red line corresponds to the giant micro-pulse profile from FD-range 72 of the 2016 dataset. Figure \ref{fig:bright} shows the scattered plot of peak flux density calculated using Equation \ref{eq:sp} for pulses with peak flux exceeding $2500$~Jy, along with pulse width at half maxima. The histogram of both peak flux density and width is plotted in the top and right panels, respectively. The bright pulses are much narrower than the averaged pulse profile, which can be found in the literature too \citep{Johnston2001}.

The pulses with a peak flux of more than $3000$~Jy and peak flux density of more than $ \simeq 1000$~Jy are found to be extracted for higher FD-ranges which cover two bright sub-components of the emission zone. The exact recording time for 2016-data is $3315$~min and $3288$~ min for 2020-data. For $1376$~MHz, the burst rate (frequency or rate at which these bursts of giant micro pulses occur) is $1/1.12$~min for bright pulses and $1/9.13$~min for giant-micro pulses from 2020-data. However, the burst rates for bright and giant-micro pulses are $1/1.68$~min and $1/25.11$~min for 2016-data, respectively. This clearly supports the theory of different activity phases of the Vela pulsar. \citet{Chen2020} had burst rate of $1/26.6$~min at $6800$~MHz and \citet{Yan2023} has reported finding $1/56$~min in L-band for giant-micro pulses.


\section{DISCUSSION}
\label{sec:discuss}

According to \citet{Kramer2002} and~\citet{Johnston2002}, the Vela pulsar has been identified as an excellent candidate for emission studies. It is due to its unique characteristics, including the close correlation between giant micro-pulses and classical giant pulses and the emission of normal pulses. The Vela pulsar serves as an ideal case for investigating the properties and mechanisms of these different types of pulsar emissions. The Vela pulsar emits pulses in a wide frequency range exhibiting various structures of individual pulses that vary from the average integrated pulses profiles ~\citep{Cairns2001, Cairns2003, Shannon2014}. These individual pulses also vary in peak flux density; hence, different nomenclature is used to differentiate them~\citep{Cordes1988, Johnston2001, Edwards2002, Palfreyman2011}. We found minimal mention of analyses on the variability of the pulse shape for the same frequency for an extensive period. In this work, the observational period lets us study and compare the Vela pulsar for the different years and peak flux density for a fixed observed frequency. The FD-integrated profile model (KD83) would reduce uncertainties arising from different individual pulse shapes because the pulse properties vary for pulses even with similar frequencies and peak flux densities.

Our fainter profiles have normal pulses. In contrast, the pulses from brighter profiles are giant micropulses that exhibit properties similar to giant pulses of the Crab pulsar and two millisecond pulsars, B1821$-$24 and B1937+21~\citep{Staelin1968, cognard1996,Romani2001}. The Vela pulsar and two other pulsars (B1706$-$44 and B0950$+$08) have been recognised as giant micro-pulse emitters~\citep{Johnston2001, Kramer2002, Johnston2002, Cairns2004}. These are examples of high-emission pulsars. Another example of such sporadic, bright emission would be PSR J0901-4624 \citep{Raithel2015}. From our analyses, we have found that the type of pulses Vela emits is connected with the FWHM of the integrated pulse profiles and is twice as many for 2020-data when the Vela pulsar is in considerably high activity mode. The two brightest pulses from 2020-data have FWHM of $0.250\,$ms and $0.305\,$ms and peak flux densities of $4989.62 \pm 0.86$~Jy and $4969.87 \pm 0.73$~Jy, making them exceptional as per the definition of \cite{Johnston2001}. The brightest pulse we found for 2016-data is from FD-range 79, with a peak flux density of $5076.25 \pm 0.09$~Jy and FWHM $0.265\,$ms.

To our knowledge, no existing literature extensively investigates these pulses together. The resultant hyperbolic nature of the FD-integrated pulse profile phase with the asymptotic value of $800$~micro-seconds could support the theory of different origins \citep{Cairns2004} for both types of pulses. The plausible hypothesis for the origin of giant pulses is the collapse of plasma-turbulent wave packets \citep{hankins2003} or emission from a rotating carousel of discharges that circulates the magnetic axis \citep{Gil2003}. Large flux densities, smaller-restricted phase windows and power-law distribution put giant and giant micro-pulses in similar categories \citep{Cairns2004}. We have also found power-law distribution for the analysed bright pulses. The comparable properties could indicate a similar emission process. The mean flux density of FD-integrated profiles exceeding a threshold of 20 was five times greater than that of the integrated profile of $\sim 18$~hr. Notably, these profiles exhibited bright pulses, which can be characterized as conforming to a power-law distribution. Furthermore, these pulses demonstrated a narrowed FWHM, measuring less than $0.5366$~ms, and their peak fluxes ranged from $1275$~Jy to $3000$~Jy.

The $89.3$~ms period of Vela pulsar may occasionally fluctuate due to activities such as glitches \citep{Palfreyman2018, Manchester2005}, bright pulses \citep{Palfreyman2011}, consecutive bright pulses \citep{Palfreyman2011}, nulling \citep{Ashton2019}. We analysed around 4425693 pulses but have not found nulling. Among these, pulses with peak flux greater than $1400$~Jy were found in the window of $0.3$~ms phase, ranging from approximately $-0.6$~ to $-0.9$~ms. From the $89.3$~ms period and $8$~ms observed radio emission region, our 59 FD-integrated profiles (FD-range 20 onwards) fall in just the $0.3$~ms phase range. This small phase region for higher FD-integrated profiles would support the theory of separate origins for normal pulses and giant micropulses \citep{Cairns2004, hankins2003, Johnston2002, Romani2001, Gil2003}.

According to the research conducted by \cite{Krishnamohan1983}, the emission of the Vela pulsar comprises four distinct emission zones, all of which we could identify in our study. Nevertheless, only two components have been found for higher FD-integrated profile 70. The smaller two emission zones were missing, which might indicate that they are too small for detection or not present. However, the other present components had different phase and peak values, which might show the emission region is moving. This difference is another point supporting the theory put forward by \cite{Cairns2004}. The Vela pulsar has classical properties such as microstructure, high degree of polarisation, S-like position angle swing, and features related to high-energy emissions, such as giant micro-pulses and different profile components~\citep{Johnston2001}. Since we can now relate giant and giant-micro pulses, the emission process can also be connected to the radio and high-energy emissions.


\section{CONCLUSION}
This study aims to extend the KD83 gated profiles of the Vela pulsar by a factor of approximately 20 in flux density and to examine the characteristics of the brighter profiles. We have also investigated the correlation between the width of the integrated pulse profiles and the emission of bright and giant micro-pulses of the Vela pulsar and found a positive correlation between them.

Our analysis indicates that the arrival times of these FD-integrated profiles have a hyperbolic relationship with their peak flux density, with an asymptote of $-0.85$~ms compared to the peak of the integrated profile at $1376$~MHz. This behaviour could suggest a connection between the disappearance of emission sub-components and their phase delay at higher peak intensities, potentially indicating that the emission region of brighter pulses is near the outermost part of the emission cone beyond which the magnetic field lines are closed. While normal pulses had four emission components, we observed that giant micro-pulses had only two components, indicating different emission regions for these two types. Moreover, we found that the emission of bright pulses follows a power-law distribution similar to that of giant micro-pulses.


\section*{ACKNOWLEDGEMENTS}

We thank the referee for the extremely helpful and positive comments, which improved the paper. The authors would like to thank Emeritus Professor John Dickey from the University of Tasmania for his feedback on the early drafts of this paper. Most of this work was based on AM's MAppSci thesis from the University of Tasmania. A.M. would like to acknowledge the financial support from the OzGrav scholarships, which made much of this work possible. S.S. acknowledges the support from Australian Government Research Training Program Scholarship. We acknowledge the use of the ATNF Pulsar Catalogue, \cite{Manchester2005} \footnote{\url{https://www.atnf.csiro.au/research/pulsar/psrcat/}}.

\section*{Data Availability}

The article's data will be shared by the corresponding author upon reasonable request.


\bibliographystyle{mnras}
\bibliography{paper} 


\bsp	
\label{lastpage}
\end{document}